\documentclass[aps,prl,twocolumn,epsfig,amsmath,amssymb]{revtex4}
\usepackage[english]{babel} \usepackage{latexsym}
\usepackage{graphicx} \usepackage{subfigure} \usepackage{epsfig}
\usepackage{psfrag}

\newcommand{\re}{\mbox{Re}\,} \newcommand{\im}{\mbox{Im}\,}

\begin{document}

\title{Stability of dark solitons in three dimensional dipolar Bose-Einstein condensates}
\author{R. Nath$^1$, P. Pedri$^{2,3}$ and L. Santos$^1$} 
\affiliation{
\mbox{$^1$Institut f\"ur Theoretische Physik , Leibniz Universit\"at
Hannover, Appelstr. 2, D-30167, Hannover, Germany}\\
\mbox{$^2$Laboratoire de Physique Théorique et Modèles Statistiques, Université Paris Sud, 91405 Orsay Cedex, France}
\mbox{$^3$Laboratoire de Physique Th\'eorique de la Mati\`ere Condens\'ee,
Universit\'e Pierre at Marie Curie,}\\
\mbox{case courier 121, 4 place Jussieu, 75252 Paris Cedex, France}\\
}

\begin{abstract}  
%
%

The dynamical stability of dark solitons in dipolar Bose-Einstein condensates is studied. 
For standard short-range interacting condensates dark solitons are unstable
against transverse excitations in two and three dimensions. On the
contrary, due to its non local character, the dipolar interaction allows
for stable 3D stationary dark solitons, opening a qualitatively novel scenario 
in nonlinear atom optics. We discuss in detail the
conditions to achieve this stability, which demand the use of an
additional optical lattice, and the stability regimes.

\end{abstract}  
\pacs{03.75.Fi,05.30.Jp} \maketitle



The physics of Bose-Einstein condensates (BECs) is, due to the 
interatomic interactions, inherently nonlinear, closely resembling 
the physics of other nonlinear systems, and in particular non linear optics. 
Non linear atom optics \cite{MeystreBook} has indeed attracted a major attention 
in the last years, including phenomena like  
four-wave mixing \cite{4WM} and condensate collapse \cite{collapse}. 
One of the major consequences of nonlinearity is the 
possibility of achieving solitons, i.e. non-dispersive self-bound solutions, 
in quasi-one dimensional BECs. Bright solitons have been reported in 
BECs with negative $s$-wave scattering length $a<0$ (the equivalent of 
self-focusing nonlinearity) \cite{Bright}. Dark solitons (DSs) have been realized 
as well in BECs with $a>0$ (self-defocusing nonlinearity) \cite{Dark}. In addition, 
optical lattices have allowed for the observation of gap solitons \cite{Gap}.


The stability of solitons depends crucially on the quasi-one dimensionality 
of the systems, which for the case of BEC solitons
demands a sufficiently strong transversal confinement of
the condensates \cite{Snake-BEC}. For the particular case of
dark solitons, if the transversal size of the system becomes comparable 
to the width of the dark soliton (typically provided by the healing length 
of the system, as discussed below), then the dark soliton plane becomes
dynamically unstable. This dynamical instability (so-called {\em snake
  instability}) has been previously studied in the context of non linear
optics \cite{Snake-nonlnr}. In the context of BEC, it has been shown that this
instability leads to a strong bending of the nodal plane, which breaks down into 
vortex rings and sound excitations \cite{Feder2000}, as experimentally observed 
in Ref.~\cite{Decay}. 


Nonlinear phenomena constitute an excellent example of the crucial role played
by interactions in quantum gases. Until recently, typical
experiments involved particles interacting dominantly via short-range 
isotropic potentials, which, due to 
the very low energies involved, are fully determined by the corresponding $s$-wave 
scattering length. However, recent experiments on cold molecules
\cite{Molecules}, Rydberg atoms \cite{Rydberg}, and atoms with large magnetic 
moment \cite{Chromium}, open a fascinating new research area, namely that 
of dipolar gases, for which the dipole-dipole interaction (DDI) plays a
significant or even dominant role. The DDI is long-range and anisotropic 
(partially attractive), and leads to fundamentally new physics in 
ultra cold gases \cite{Dipoles}. 
Time-of-flight experiments in Chromium have allowed for the first
observation of DDI effects in cold gases \cite{Expansion}, 
which have been remarkably enhanced recently by means of Feshbach resonances \cite{Feshbach}.


Dipolar gases present a rich non-linear physics, since the DDI leads 
to non-local non-linearity, similar as that encountered in plasmas \cite{Plasma}, 
nematic liquid crystals \cite{Nematics}, thermo-optical materials \cite{ThermRef} 
and photorrefractive crystals \cite{Photoref}. Nonlocality leads to a wealth of novel phenomena in nonlinear 
physics, as the modification of modulation instability \cite{ModInst},
the change of the soliton interaction \cite{SolitInt}, and the stabilization of azimuthons \cite{Azimuthon}. 
Particularly interesting is the possibility of stabilization of localized
waves in cubic nonlinear materials with a symmetric 
nonlocal nonlinear response \cite{Bang}. 
Multidimensional solitons have been experimentally 
observed in nematic liquid crystals \cite{Nematics-Soliton} and in
photorrefractive screening solitons \cite{spiraling}. 
Recently we showed that under realistic conditions, 
2D bright solitons may be generated in dipolar BEC \cite{Pedri05}.  
Interestingly, due to the non-local nature of the DDI, the soliton-soliton
scattering presents strong inelastic resonances, as well
as spiraling motion \cite{Nath}, similar to that observed in photorrefractive 
materials \cite{spiraling}.


In this letter, we show that the long-range character of the DDI may have 
striking consequences for the stability of dark solitons in dipolar BECs. 
Contrary to usual BECs, for which, as mentioned above, dark nodal planes
become unstable when departing from the one-dimensional condition, the DDI 
may stabilize dark nodal planes even if the transversal size of the condensate
becomes arbitrarily much wider than the condensate healing length. This
stabilization is purely due to the long-range character of the DDI. We study
in detail the conditions for this stabilization, and the stabilization regimes.



In the following, we consider a dipolar BEC of particles with mass $m$ and 
electric dipole $d$ (the results are equally valid for magnetic dipoles) 
oriented in the $z$-direction by a sufficiently large external field, and 
that hence interact via a dipole-dipole potential: 
$V_d(\vec{r})= \alpha d^2 (1-3\cos^2(\theta))/r^3$, where $\theta$ 
is the angle formed by the vector joining the interacting particles 
and the dipole interaction. The coefficient $\alpha$ can be tuned within 
the range $-1/2\leq\alpha\leq 1$ by rotating the external field that 
orients the dipoles much faster than any other relevant time scale 
in the system \cite{Tuning}. At sufficiently low temperatures the physics 
of the dipolar BEC is provided by a non-local non-linear Schr\"odinger 
equation (NLSE) of the form:
\begin{eqnarray}
&&i\hbar\frac{\partial}{\partial t}\Psi(\vec r,t)=
\left [ 
-\frac{\hbar^2}{2m}\nabla^2+V_{ol}(x,y)
+g|\Psi(\vec r,t)|^2 \right\delimiter 0 \nonumber \\
&&+ \left\delimiter 0  \int d\vec r' V_d(\vec r-\vec r')
|\Psi(\vec r',t)|^2
\right ]\Psi(\vec r,t),
\label{GPE}
\end{eqnarray}
where $g=4\pi\hbar^2a/m$, with $a$ the $s$-wave scattering length 
(we consider $a>0$) and $m$ the particle mass. For reasons that will become clear below, the BEC is
assumed to be in a 2D optical lattice, 
$V_{ol}(x,y)=sE_R(\sin^2(q_{l}x)+\sin^2(q_{l}y))$, where $E_R=\hbar^2q_l^2/2m$ 
is the recoil energy, $q_l$ is the laser wave vector and $s$ is a dimensionless 
parameter providing the lattice depth. In the tight-binding regime 
(i.e. for a sufficiently strong lattice but still maintaining coherence), 
we may write $\Psi(\vec r,t)=\Sigma_{i,j}f_{ij}(x,y)\psi_{i,j}(z,t)$, where 
$f_{ij}(x,y)$ is the Wannier function associated to the 
lowest energy band and the site located at $(bi,bj)$, with $b=\pi/q_l$. 
Substituting this ansatz in Eq.~(\ref{GPE}) we 
obtain a discrete NLSE \cite{Smerzi}. We may then return to a continuous equation, 
where the lattice is taken into account in an effective mass along 
the lattice directions and in the renormalization of the coupling constant \cite{Lattice}.
\begin{eqnarray}
&&i\hbar\frac{\partial}{\partial t}\Psi(\vec r,t)=
\left [ 
-\frac{\hbar^2}{2m^*}\nabla_\perp^2-\frac{\hbar^2}{2m}\frac{\partial^2}{\partial z^2}
+\tilde{g}|\Psi(\vec r,t)|^2 \right\delimiter 0 \nonumber \\
&&+ \left\delimiter 0  \int d\vec r' V_d(\vec r-\vec r')
|\Psi(\vec r',t)|^2
\right ]\Psi(\vec r,t),
\label{GPE1}
\end{eqnarray}
where $\tilde g=b^2g\int f(x,y)^4dxdy+g_dC$ \cite{coefc}, with $g_d=\alpha 8\pi d^2/3$, 
$m^*=\hbar^2/2b^2J$ is the effective mass, and 
$J=\int dx dy f_{ij}(x,y)[-(\hbar^2/2m)\nabla^2_\perp+V_{ol}(x,y)]f_{i'j'}(x,y)$,
for neighboring sites $(i,j)$ and $(i',j')$.
The validity of Eq.~(\ref{GPE1}) is limited to radial momenta $k_\rho\ll 2\pi/b$, 
in which one can ignore the discreteness of lattice. 
In the following we use the convenient dimensionless parameter $\beta=g_d/\tilde g$, 
that characterizes the strength of the DDI compared to the short range
interaction. The Fourier transform of the DDI: 
$V_d(\vec k)=g_d[3\cos^2\theta_k-1]/2$, with $\cos^2\theta_k=k_z^2/|\vec k|^2,$ 
is needed later for the calculations.



Due to its partially attractive character, the stability of a dipolar BEC is a
matter of obvious concern \cite{Dipoles}. A Bogoliubov analysis of an 
homogeneous dipolar condensate reveals that the dispersion relation for
quasiparticles is of the form 
$\epsilon (\vec k)=[E_{kin}(\vec k)[E_{kin}(\vec k)+E_{int}(\vec k)]]^{1/2}$, 
where $E_{kin}(\vec k)=\hbar^2k_{\rho}^2/2m^*+\hbar^2k_{z}^2/2m$ is 
the kinetic energy, and $E_{int}(\vec k)=2(g+\tilde V_d(\vec k))n_0$ 
is the interaction energy, which includes both short-range and dipolar parts. 
Note that $V_d(\vec k)$ may be positive or negative, and hence for low momenta
(phonon excitations) the dynamical instability (phonon instability) is just
prevented if $-1<\beta<2$. If $g_d>0$, phonons with $\vec k$ lying on the $xy$ plane are 
unstable if $\beta > 2$, while for $g_d<0$ phonons with $\vec k$ along $z$ 
are unstable if $\beta < -1$ \cite{Vortex}.



In this paper, we are particularly concerned about the stability of a dark soliton in 
a three dimensional dipolar BEC. We assume that the dark soliton lies on the
$xy$ plane, hence the solution can be written in the following form: $\Psi_0(\vec r,t)=\psi_0(z)\exp[-i\mu t/\hbar]$, where $\mu$ is the chemical potential. Introducing this 
expression into Eq.~(\ref{GPE}) we obtain a one-dimensional NLSE in $z$ of the form:
\begin{eqnarray}
&&\mu\psi_0(z)=
\left [ 
-\frac{\hbar^2}{2m}\frac{\partial^2}{\partial z^2}
+\bar{g}|\psi_0(z)|^2\right ]\psi_0(z).
\label{GPE2}
\end{eqnarray}
Note that, since $\psi_0$ is independent of $x$ and $y$, 
in this equation the DDI interaction just regularizes the value of 
the local coupling constant $\bar g=\tilde g+g_d$. This equation allows 
for a simple solution describing a dark soliton, $\psi_0=\sqrt{n_0}\tanh(z/\zeta)$, 
where $\zeta=\hbar/\sqrt{m\bar gn_0}$ is the corresponding healing length and $n_0$ is the bulk density. 
Note that, interestingly, due to the modification of the local coupling constant
the size $\zeta$ of the dark soliton depends on the DDI.



We are especially interested in the dynamical stability of these nodal planes.
To study this stability we perform a Bogoliubov analysis, considering a
transversal perturbation of the dark soliton planes:  
$\Psi(\vec r,t)=\Psi_0(\vec r,t)+\chi(\vec r,t)\exp (-i\mu t/\hbar)$, 
where $\chi (\vec r,t)=u(z)\exp[i(qx-\epsilon
t/\hbar)]+v(z)^*\exp[-i(qx-\epsilon^*t/\hbar)]$, 
where $q$ is the momentum of the transverse modes with energy $\epsilon$. 
Introducing this ansatz into Eq.(\ref{GPE}) and linearizing in $\chi$, 
one obtains the corresponding Bogoliubov-de Gennes (BdG) equations 
for the excitation energies $\epsilon$ and the corresponding eigenfunctions  
$f_\pm=u\pm v$:
\begin{eqnarray}
&&\epsilon f_-(z)=
\left [ 
-\frac{\hbar^2}{2m}\left (\frac{\partial^2}{\partial z^2}-\frac{m}{m^*}q^2 \right )-\mu
+3\bar{g}\psi_0(z)^2\right ] f_+(z) \nonumber \\
&&-\frac{3}{2}g_dq\psi_0(z) \int_{-\infty}^{\infty} dz' \exp(-q|z-z'|)\psi_0(z')f_+(z'),
\label{BdG1}
\end{eqnarray}
\begin{eqnarray}
\epsilon f_+(z)=
\left [ 
-\frac{\hbar^2}{2m}\left (\frac{\partial^2}{\partial z^2}-\frac{m}{m^*}q^2 \right )-\mu
+\bar{g}\psi_0(z)^2\right ] f_-(z) 
\label{BdG2}
\end{eqnarray}

Note that the dipole interaction has two main effects. On one side, as
mentioned above, it leads to a regularized $\bar g$. On the other side, 
it introduces a qualitatively new term in the second line of Eq.~\ref{BdG1}. 
Whereas the first effect just leads to a quantitative modification of the dark
soliton width, the second effect is a purely dipole-induced non local effect,
which, as we show below, may lead to remarkable consequences for the dark
soliton stability. For every transverse momentum $q$ we determine 
the lowest eigenenergy, which provides the dispersion law $\epsilon(q)$. 

\begin{figure} 
\begin{center}
\psfrag{x}{$q\zeta$}
\psfrag{y}{\hspace{-1cm}{$\displaystyle\frac{\im(\epsilon)}{\mu}$}}
\includegraphics[width=6.0cm]{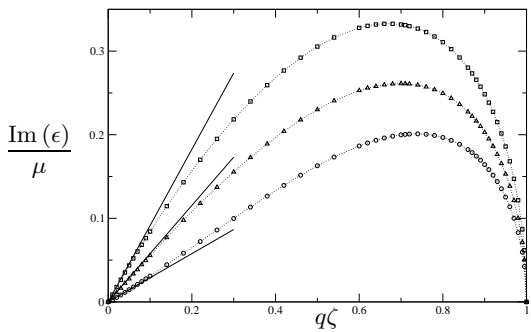}
\end{center} 
\caption{Imaginary part of the excitation energies of a DS
for $m/m^*=1$, and $\beta=0$ (triangles), $-0.5$ (squares) 
and $1$ (circles). Solid lines correspond to the analytical result for low momenta.
The real part of the excitation energies is equal to zero for the range of momenta considered in the figure.}  
\label{fig:1}
\vspace*{-0.2cm}  
\end{figure}

When $\beta=0$ (no DDI) and $m/m^*=1$ (no lattice), we recover the same 
BdG equations discussed in the context of standard short-range interacting condensates
\cite{Snake-BEC}. It has been shown that in that case the dispersion law 
$\epsilon(q)$ is purely imaginary for $q\zeta<1$ (Fig.~\ref{fig:1}). Hence 
dark-soliton planes in homogeneous three-dimensional short-range interacting
condensates, are dynamically unstable against transverse modulations. This 
so-called {\em snake instability} has been experimentally observed in 
non linear optics \cite{Snake-nonlnr} and recently in the context of BEC
\cite{Decay}. In the latter case, the 
dark-soliton plane bends and decays into more stable structures such as 
vortex rings and sound excitations \cite{Feder2000}. The stabilization of the
soliton demands a strong radial 
confinement \cite{Snake-BEC}, i.e. 
the dark-solitons cannot be considered any longer as three-dimensional, 
but on the contrary acquire a one-dimensional character.

\begin{figure} 
\begin{center}
\psfrag{x}{\hspace{0.5cm}$\beta$}
\psfrag{y}{\hspace{-0.5cm}{$\displaystyle\frac{m}{m^*}$}}
\includegraphics[width=6.0cm]{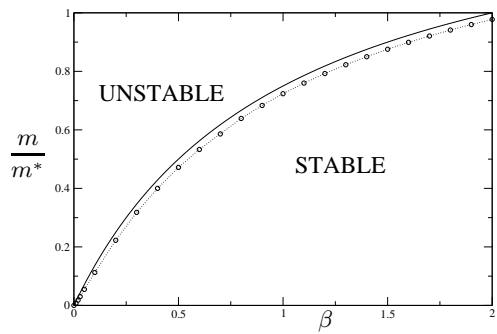}
\end{center} 
\caption{Analytical (solid line) and numerical (empty circles) results for the
  stable/unstable phonon regimes as a function of two paremeters $\beta$ and $m/m^*$.}  
\label{fig:2}
\end{figure}

\begin{figure} 
\begin{center}
\psfrag{x}{$q\zeta$}
\psfrag{y}{\hspace{-1cm} $\displaystyle\frac{\re(\epsilon)}{\mu}$}
\includegraphics[width=7.5cm]{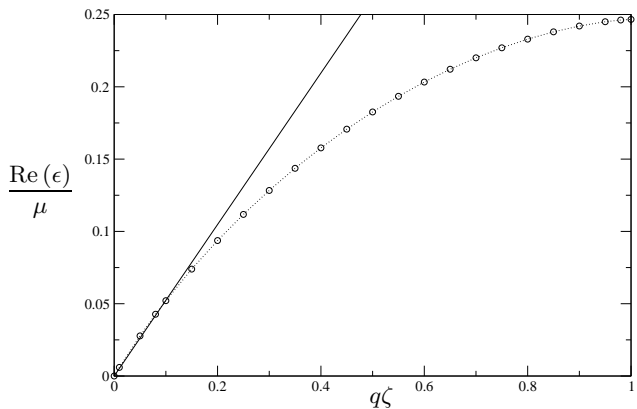}
\end{center} 
\caption{Real part of the excitation energies of a DS
for $m/m^*=.1$, and $\beta=1.6$. Solid line corresponds to the analytical result for low momenta while empty circles correspond to numerical results.
The imaginary part of the excitation energies is equal to zero for the range of momenta considered in the figure.}  
\label{fig:3}
\end{figure}

In the presence of DDI ($\beta\neq 0$) but without lattice  
($m/m^*=1$), the transverse instability persists, 
since $\epsilon(q)$ remains purely imaginary for $q\zeta<1$. 
For $\beta>0$ ($\beta<0$), $|\epsilon(q)|$ decreases (increases) 
when $|\beta|$ grows (Fig.~\ref{fig:1}). The situation
changes completely when $m/m^*$ is sufficiently small since, as discussed 
below, $\epsilon(q)$ is purely real and, as a consequence,
the dark soliton becomes dynamically stable (see Fig.\ref{fig:3}). 
As we detail in the following, this remarkable fact can be clearly 
understood by means of a transparent physical picture.
First we notice that for low momenta the spectrum 
is always linear in $q$, suggesting the idea
that for low momenta the system may be described by an elastic model 
of the two dimensional nodal plane of the dark soliton. 
The Lagrangian density for the nodal plane reads
 \begin{eqnarray}
{\cal L}\left(\frac{d\phi}{dt},\vec{\nabla} \phi\right)=\frac{M}{2}\left ( \frac{d\phi}{dt} \right )^2-\frac{\sigma}{2}|\vec{\nabla} \phi|^2
\label{LAGR}
\end{eqnarray}
where $\phi$ is the displacement field of the nodal plane 
from the ground state, $M$ is the mass for unit area and $\sigma$ 
plays the role of a surface tension. 
The mass $M$ of the soliton can be easily 
calculated expanding the energy of a moving soliton 
up to second order in the velocity. We obtain
$M=-4\hbar n_0/c$, where $c=\sqrt{\bar{g}n_0/m}$ is the 
sound velocity. Notice that $M<0$ since the dark soliton 
represents an absence of atoms. The surface tension can 
be calculated inserting a suitable variational ansatz 
$\Psi_{\rm var}(\vec r)=\sqrt{n_0}
\tanh ((z-\sqrt 2\alpha\cos (qx))/\zeta)$ (describing   
a transverse modulation of the nodal plane with amplitude  
$\alpha$ and momentum $q$) in the energy functional 
and expanding up to second order in $\alpha$ and $q$: 
\begin{eqnarray}
\sigma=\frac{4n_0\hbar^2}{3\zeta m^*}-2g_dn_0^2\zeta
\label{SIGMA}
\end{eqnarray}
This expression can be considered as one of the main results of this Letter. 
The nature of the eigenmodes $\omega^{2}=(\sigma/M)q^{2}$ crucially 
depends on the sign of $\sigma/M$. In the absence of 
DDI ($\beta=0$), $\sigma$ is always positive and hence
the modes are purely imaginary. The dark soliton is dynamically unstable
against the above mentioned snake instability. Note that for $\beta=0$ and $m/m^*=1$
our result coincides with the one found in Ref.~\cite{Snake-BEC}. 
In the absence of an additional optical lattice, 
the dynamical instability of the DS at low $q$ dissapears for $\beta>2$, i.e. 
for situations for which the homogeneous 
dipolar BEC as a whole is itself, as commented above, unstable against local collapses. 
Increasing the depth of the lattice potential 
reduces the role of the kinetic energy term $(m/m^*)q^2$ in Eqs.~(\ref{BdG1}) and 
(\ref{BdG2}) (or equivalently reduces the first term in Eq.~(\ref{SIGMA})) 
and hence enhances the role of the DDI. A sufficiently large DDI or small $m/m^*$ such that 
\begin{equation}
\frac{m}{m^*}<\frac{3\beta}{2(1+\beta)}
\end{equation}
leads to stable low-energy phonons with $q\rightarrow 0$. 
We have evaluated from a direct numerical calculation of the BdG equations
(\ref{BdG1}) and (\ref{BdG2}) the stability threshold at which
$\epsilon(q\rightarrow 0)$ becomes real. Fig.~\ref{fig:2} compares our 
numerical and analytical results, which are in excellent agreement.

When $m/m^*$ decreases further or $\beta$ grows, a wider regime of
low momenta is stabilized (Fig.~\ref{fig:3}). Note in 
Fig.~\ref{fig:3} that the dispersion law at low momenta is very accurately 
described by our analytical results. 
A sufficiently strong optical lattice and large DDI 
can stabilize all the modes with momenta up to $q\sqrt{m/m^*}\zeta\sim 1$. 
Although we observe instabilities for momenta $q\sqrt{m/m^*}\zeta\sim 1$, 
this large-momentum instability is typically irrelevant, 
since for sufficiently small $m/m^*$ it concerns
momenta much larger than the lattice momentum. Although our
effective mass theory breaks down for such momenta, it becomes clear that 
such high momentum instabilities are physically prevented by the 
zero point oscillations at each lattice site.  



Summarizing, contrary to short-range interacting BECs, 
where stable dark solitons demand a sufficiently 
strong transverse confinement, dipolar BECs allow for stable 
dark solitons of arbitrarily large transversal sizes. 
We have obtained the stability conditions, which demand a sufficiently 
large dipole and a sufficiently deep optical lattice in the nodal plane.
We presented analytical results for the lowest part of the spectrum which 
agree with those obtained numerically. 
The stabilization of nodal planes 
is purely linked to the long-range nature of the DDI,  
opening a qualitatively new scenario in 
non linear atom optics.

\acknowledgements
We acknowledge fruitful discussion with Lev P. Pitaevskii, L. Pricoupenko and G. V. Shlyapnikov. 
This work was supported by the DFG (SFB-TR21, SFB407, SPP1116), by the Minist\`ere de la
Recherche (grant ACI Nanoscience 201), by the ANR (Grants Nos. NT05-2\_42103 and 05-Nano-008-02), and by the IFRAF Institute.
 LPTMC is UMR 7600 of CNRS.


\begin{thebibliography}{99}
\bibitem{coefc} $C\simeq\sum_{i\neq 0,j\neq 0}|\tilde{n}(2\pi i/b,2\pi j/b)|^2$ where $\tilde{n}$ is the Fourier transform of $n$ and   
$n(x,y)=|f(x,y)|^2$.
\end{thebibliography}
\end{document}